# Spin and Lattice Structure of Single Crystal $SrFe_2As_2$

Jun Zhao[1], W. Ratcliff II[2], J. W. Lynn[2], G. F. Chen[3], J. L. Luo[3], N. L. Wang[3], Jiangping Hu[4], and Pengcheng Dai[1,5,*]

[1] Department of Physics and Astronomy, The University of Tennessee, Knoxville, Tennessee 37996-1200, USA

[2] NIST Center for Neutron Research, National Institute of Standards and Technology, Gaithersburg, Maryland 20899-6012 USA

[3] Beijing National Laboratory for Condensed Matter Physics, Institute of Physics, Chinese Academy of Sciences, Beijing 100080, China

[4]Department of Physics, Purdue University, West Lafayette, Indiana 47907, USA

[5]Neutron Scattering Science Division, Oak Ridge National Laboratory, Oak Ridge, Tennessee 37831, USA

## Abstract

We use neutron scattering to study the spin and lattice structure on single crystals of $SrFe_2As_2$, the parent compound of the FeAs based superconductor $(Sr,K)Fe_2As_2$. We find that $SrFe_2As_2$ exhibits an abrupt structural phase transitions at 220K, where the structure changes from tetragonal with lattice parameters $c > a = b$ to orthorhombic with $c > a > b$. At almost the same temperature, Fe spins in $SrFe_2As_2$ develop a collinear antiferromagnetic structure along the orthorhombic *a*-axis with spin direction parallel to this *a*-axis. These results are consistent with earlier work on the *R*FeAsO (*R* = rare earth elements) families of materials and on $BaFe_2As_2$, and therefore suggest that static antiferromagnetic order is ubiquitous for the parent compound of these FeAs-based high-transition temperature superconductors.

Understanding the structural, electronic, and magnetic properties of parent compounds of high-transition temperature (high-$T_c$) superconductors is an essential step in developing a microscopic theory for superconductivity. For high-$T_c$ copper oxides, the parent compounds are antiferromagnetic (AFM) Mott insulators, where the nearest neighbor $Cu^{2+}$ spins in the $CuO_2$ plane arrange themselves antiferromagnetically[1]. In the case of the newly discovered Fe-As based high-$T_c$ superconductors[2-8], while the Fe ions in parent compounds LaFeAsO (ref. 9,10), CeFeAsO (ref. 11), and $BaFe_2As_2$ (refs. 12,13) are found to exhibit commensurate static AFM long range order, there have been reports of the iron ordering being absent in NdFeAsO (refs. 14, 15). Therefore it is still unclear whether the observed Fe AFM order is a universal feature for the parent compounds of these Fe-As based superconductors. Furthermore, all previous neutron scattering experiments on these FeAs-based materials were carried out on polycrystalline samples[9,10,12,13,14], where it was not possible to determine the spin direction, or in most cases the AFM ordering wave vector.

In this paper, we report single crystal neutron scattering studies of the structural and magnetic phase transitions for $SrFe_2As_2$, the parent compound of the $(Sr,K)Fe_2As_2$ superconductors[8,16]. Previous transport, $^{57}$Fe Mössbauer, and X-ray diffraction experiments[17,18,19] have shown that $SrFe_2As_2$ exhibits structural and magnetic phase transitions at 203 K, where the crystal structure changes abruptly from tetragonal (*I4/mmm*) to orthorhombic (*Fmmm*). Our neutron scattering experiments confirm the findings of the x-ray measurements for the structural transition, while we are able to determine conclusively that the Fe spins in $SrFe_2As_2$ order antiferromagnetically along the orthorhombic *a*-axis and ferromagnetically along the *b*-axis, with the moment direction along the *a*-axis (Figs. 1a and 1b). These measurements, together with the recent discovery of static long-range AFM order of the Fe sublattice in NdFeAsO (ref. 20), suggest that the collinear AFM order shown in Fig. 1b is ubiquitous for the parent compounds of the FeAs-based superconductors.

We use neutron diffraction to study the structural and magnetic phase transitions in crystals grown using the method described in Ref. 21. Our experiments were carried out on the conventional triple-axis spectrometer BT-9 at the NIST Center for Neutron Research, Gaithersburg, Maryland. The neutron wavelength employed was 2.359 Å using a pyrolytic graphite (PG) monochromator, and PG filter to suppress higher-order reflections to achieve a monochromatic incident beam. The collimations were 40′-47′-S-40′-80′. We denote positions in momentum space using $Q = (H, K, L)$ in reciprocal lattice units (r.l.u.) in which Q (in $Å^{-1}$) = $(H2\pi/a, K2\pi/b, L2\pi/c)$, where $a = 5.5695(9)$, $b = 5.512(1)$, $c = 12.298(1)$ Å are lattice parameters in the orthorhombic state at 150 K. The sample (~5×5×0.5 $mm^3$, mosaic ~ 0.3°) was mounted on an aluminum plate and aligned in the [*H*, 0, *L*] zone inside a sealed aluminum container with helium exchange gas and mounted on the cold finger of a closed cycle helium refrigerator.

Figures 1a and 1b summarize our experiments, which show the Fe spin arrangements with respect to the orthorhombic low temperature crystal structure. To obtain integrated magnetic intensities necessary for comparison with magnetic structure factor calculations, we carried out radial (θ:2θ) as well as rocking (θ) scans for a series of magnetic (1, 0, *L*) and (3, 0, *L*) peaks, where $L = 1, 3, 5, \ldots$ Figures 1c and 1d show scans for the (1,0,1) and (1,0,3) peaks below and above the AFM ordering temperature. Sharp, resolution-limited magnetic peaks are observed at 10 K, and completely disappear at 250 K, consistent with establishment of long-range AFM order. A detailed investigation of the low-temperature magnetic Bragg peaks in the (*H*, 0, *L*) zone revealed an ordered magnetic structure of Fe ions consistent with previous results on LaFeAsO (ref. 9), CeFeAsO (ref. 11), NdFeAsO (ref. 20) and $BaFe_2As_2$ (ref. 13).

In previous X-ray and neutron diffraction work on $BaFe_2As_2$ and $SrFe_2As_2$, it was found that structural distortion occurs almost simultaneously with AFM order[12,13,16-19]. To confirm this in our single crystal of $SrFe_2As_2$, we carried out neutron diffraction measurements focusing on the $(2,2,0)_T$ nuclear Bragg peak, where T denotes the high temperature tetragonal phase. As a function of decreasing temperature, the $(2,2,0)_T$ peak abruptly splits into the $(4, 0, 0)_O$ and $(0, 4, 0)_O$ Bragg peaks below 220 ± 1 K as shown in Fig. 2. Here the subscript O denotes orthorhombic symmetry and the observation of both $(4, 0, 0)_O$ and $(0, 4, 0)_O$ peaks indicates that our single crystal has equally populated twin domains in the orthorhombic phase. Figure 2b compares the structural phase transition and magnetic order parameter in detail as a function of temperature. It is evident that the structural transition occurs more abruptly compared to the magnetic phase transition. By normalizing magnetic peaks with nuclear structural peaks using the magnetic structure shown in Figs. 1a and 1b, we estimate that the ground state ordered iron moment is approximately 0.94(4) $\mu_B$ at 10 K, where numbers in parentheses indicate one standard deviation statistical uncertainty in the last decimal place and $\mu_B$ denotes Bohr magneton.

In previous neutron diffraction work on powder samples of LaFeAsO (ref. 9), CeFeAsO (ref. 11), NdFeAsO (ref. 20) and $BaFe_2As_2$ (ref. 13), it was found that the Fe spins order antiferromagnetically along one axis of the low-temperature orthorhombic structure and ferromagnetically along the other axis. However, the actual AFM and ferromagnetic ordering directions, as well as the Fe moment direction, were not determined. To determine the direction of the AFM ordering in $SrFe_2As_2$, we carefully probed the (3, 0, 3) magnetic Bragg reflection. Fig. 3a shows a radial scan for the magnetic scattering, where we only observed a single (magnetic) peak. Removing the PG filter allows both (6, 0, 6) and (0,6,6) orthorhombic nuclear Bragg peaks to be observed via $\lambda/2$ in the incident beam. We see that the magnetic peak corresponds to the smaller diffraction angle, which establishes that the AFM ordering is along the *a*-axis. A further check is provided in Fig. 3b, where the diffraction angle was set to the higher angle reflection and rocking curves were performed, with and without the PG filter. The only peak observed is the (0,6,6) nuclear reflection, when the filter was removed. This demonstrates that the only magnetic reflection is the (3,0,3) peak. Therefore, our experiments conclusively identify the AFM ordering direction as along the long *a*-axis direction of the orthorhombic $SrFe_2As_2$ unit cell.

To determine the Fe moment direction, we carried out integrated intensity measurements for a serious of magnetic Bragg reflections. Group symmetry analysis performed in the low temperature *Fmmm* phase restricts the moments to be either along the *a*-axis, *b*-axis, or *c*-axis, assuming that the magnetic transition is second order; if it is first order then there are no restrictions on the spin direction. However, if the dominant interactions are determined by Fe-As-Fe exchange, then the orthorhombic structure dictates that the diagonal exchange $J_2$ in Fig. 1b should be the same for both diagonal directions since the bond angles and distances are identical. Therefore we expect the Fe spins in $SrFe_2As_2$ to point either along the *a*-axis or along the *b*-axis, and this is indeed the case. Assuming the Fe spin direction is $\phi$ away from the *a*-axis (inset in Fig. 4a), the least square fit of our magnetic structure factor calculations indicates excellent agreement with a $\chi^2=3.8$ for moment along the *a*-axis ($\phi = 4 \pm 3°$). Therefore, it is clear that the moment direction is along *a*, and the spin structure is as shown in Figs. 1a and 1b.

The collinear antiferromagnetic structure can be described in an effective $J_1$ -$J_2$ – $J_z$ Heisenberg model[22-24], where $J_1$ and $J_2$ are the antiferromagnetic exchange couplings between the nearest neighbor and second nearest neighbor Fe atoms, respectively, and $J_z$ is the exchange coupling between FeAs layers (Fig. 1b). When $J_1 < 2J_2$, the model has a collinear antiferromagnetic ground state and also an Ising nematic order state at high temperature which can couple to the structure transition in the general Ginzburg-Landau approach. When $J_z/ J_2$ is larger than 0.005, the collinear magnetic and Ising nematic transition temperatures are very close[23]. In this model, by including the coupling between the Ising order and the lattice, the structure transition is expected to happen at the same transition temperature as that of the collinear magnetic transition, which is a good description of the current case in $SrFe_2As_2$, where the inter-layer coupling is much larger than that in RFeAsO compounds.

The observed configuration of orthorhombic lattice distortion and the corresponding spin arrangement reveals that the nearest neighbor AFM exchange coupling $J_1$ is not a simple result of a superexchange interaction arising from electron hopping through the As ion, since this requires $J_{1a} > J_1 > J_{1b}$ after the lattice distortion in order to save total energy[22]. The fact that the ferromagnetic exchange is along the short ($b$) axis of the orthorhombic structure suggests the presence of a significant direct ferromagnetic exchange coupling. That is, $J_1 = J_1^{\,s} - J_1^{\,d}$ where $J_1^{\,s}$ is the superexchange AFM contribution and $J_1^{\,d}$ is the ferromagnetic part from direct Fe-Fe exchange.

The small lattice distortion has little effect on the local onsite energy, but can directly change the electron hopping amplitude. Assume that the high-temperature in-plane tetragonal lattice constant $a_T$ is split into orthorhombic $a = a_T + \delta$ and $b = a_T - \delta$. Then the change of the distance between Fe and the nearest neighbor As is given by the leading order $\delta^2/4$ (Fig. 1b). Therefore, the change in the hopping amplitude $t'$ after lattice distortion is $\Delta t' \propto \delta^2$. Since the superexchange $J_1^{\,s} \propto t'^4$, its changes after the lattice distortion should be $\Delta J_1^{\,s} \propto -\delta^2$. This means that the reduction in $J_1^{\,s}$ is a second order effect of the lattice distortion. On the other hand, since the direct ferromagnetic exchange $J_1^{\,d}$ is proportional to the hopping amplitude $t$, the leading order changes of $J_1^{\,d}$ along the $a$- and $b$-axes after the lattice distortion should be $\Delta J_{1a}^{\,d} \propto -\delta$ and $\Delta J_{1b}^{\,d} \propto \delta$, respectively. Therefore, $J_1$ increases along the $a$-axis and decreases along the $b$-axis after the lattice distortion.

In FeAs-based materials, the AFM order is rapidly suppressed upon doping. This immediately suggests a decrease of $J_{1a}$ with increasing doping. Since the decrease of $J_{1a}$ mostly arises from the increase of the direct exchange $J_{1a}^{d}$, the long (*a*) axis of the orthorhombic structure is expected to be suppressed with increasing electron or hole doping. This phenomenon has indeed been observed in electron-doped CeFeAsO, where the long (*a*) axis of the orthorhombic structure is reduced upon doping F while the short (*b*) axis is unaffected (see Fig.3 of ref.11). Therefore, to obtain a comprehensive understanding of the mechanism of superconductivity in these FeAs-based superconductors, one must consider both the direct- and super-exchange interactions and their relationship to lattice distortion effects.

In summary, we have determined the AFM ordering wave vector and spin direction in $SrFe_2As_2$, the parent compound of the $(Sr,K)Fe_2As_2$ superconductors. Since recent neutron scattering[20] and μSR experiments[25] also independently confirmed that the Fe spins in NdFeAsO orders antiferromagnetically with the same spin structure as LaFeAsO (ref. 9), CeFeAsO (ref. 11), and $BaSr_2As_2$ (ref. 13), it is safe to assume that the collinear static antiferromagnetic structure shown in Fig. 1b is ubiquitous for the parent compound of these FeAs-based high-$T_c$ superconductors.

Note added: Upon finishing the present paper, we became aware of a neutron powder diffraction work which also concluded that Fe spins in $SrFe_2As_2$ order antiferromagnetically in the *a*-axis direction with ordered moment along the *a*-axis. However, these authors did not show how such conclusion was reached from their data[26].

**References:**

*e-mail: daip@ornl.gov

**Acknowledgements** This work is supported by the US National Science Foundation through DMR-0756568, by the US Department of Energy, Division of Materials Science, Basic Energy Sciences, through DOE DE-FG02-05ER46202. This work is also supported in part by the US Department of Energy, Division of Scientific User Facilities, Basic Energy Sciences. The work at the Institute of Physics, Chinese Academy of Sciences, is supported by the National Science Foundation of China, the Chinese Academy of Sciences and the Ministry of Science and Technology of China.

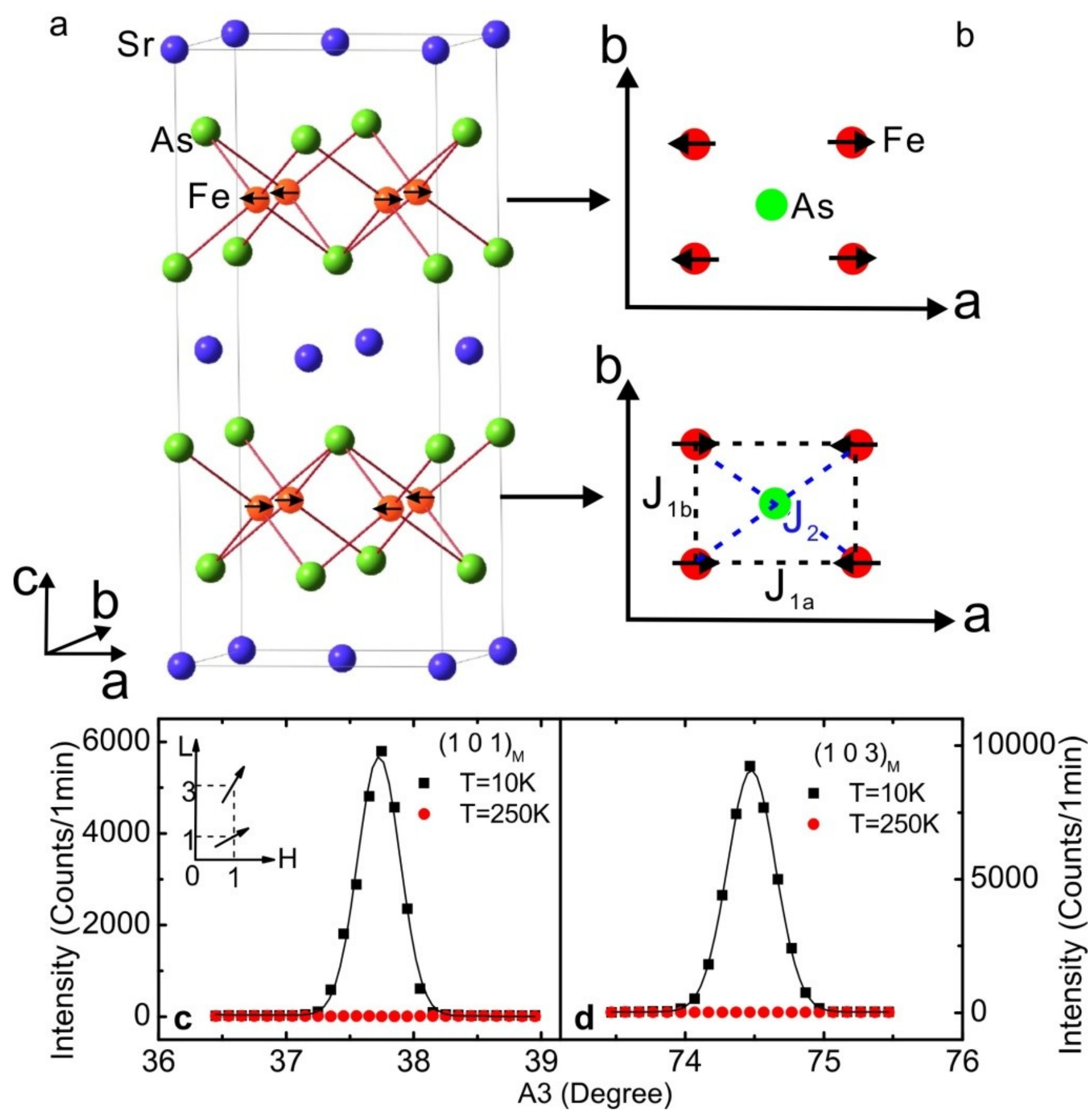

Figure 1 (color online) Crystal and magnetic structures of $SrFe_2As_2$. **a**) The three dimensional antiferromagnetic structures of Fe in $SrFe_2As_2$ as determined from our neutron diffraction data. **b**) The in-plane magnetic structure of Fe in the orthorhombic unit cell of $SrFe_2As_2$. The Fe moments are along the *a*-axis and form an AFM collinear structure along the *a*-axis direction and ferromagnetic along the *b*-axis direction, while nearest-neighbor Fe spins along the *c*-axis are anti-parallel, identical to that of LaFeAsO (ref. 9),. $J_{1a}$, $J_{1b}$, and $J_2$ indicate the effective exchange couplings. **c,d**) Radial scans through the magnetic (1, 0, 1) and (1, 0, 3) magnetic Bragg peaks below and above the Néel temperature, showing clear resolution-limited magnetic peaks.

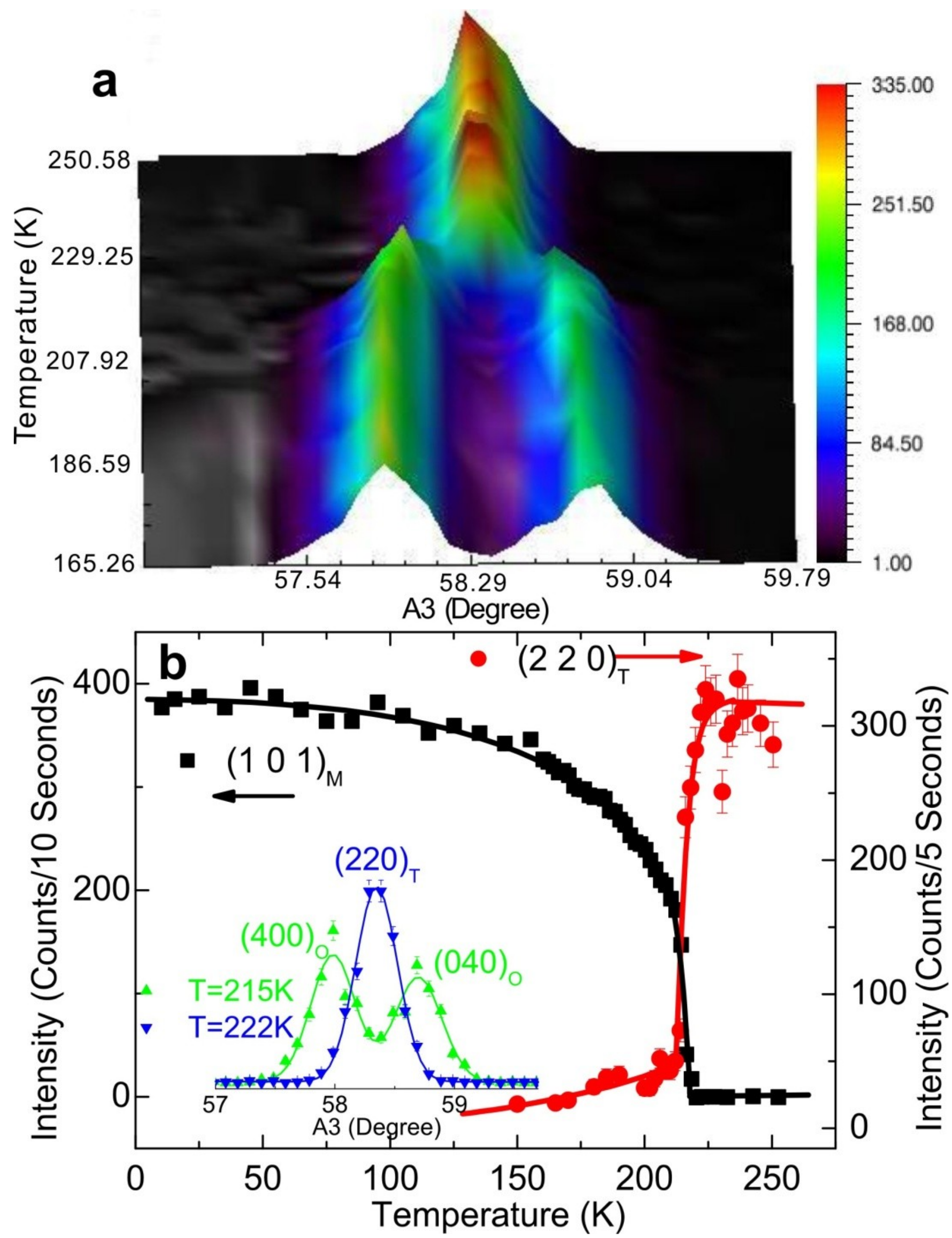

Figure 2 (color online) Structural and magnetic phase transition as a function of temperature in single crystal $SrFe_2As_2$. **a**) Temperature dependence of the $(2, 2, 0)_T$ structural peak, showing that it abruptly splits below 220 ± 1 K. The data were collected using 10′-10.7′-S-10′-80′ collimation. **b**) Comparison of structural distortion and magnetic order parameter, both occurring at essentially the same temperature. The structural transition is first order, while the magnetic transition appears to be continuous.

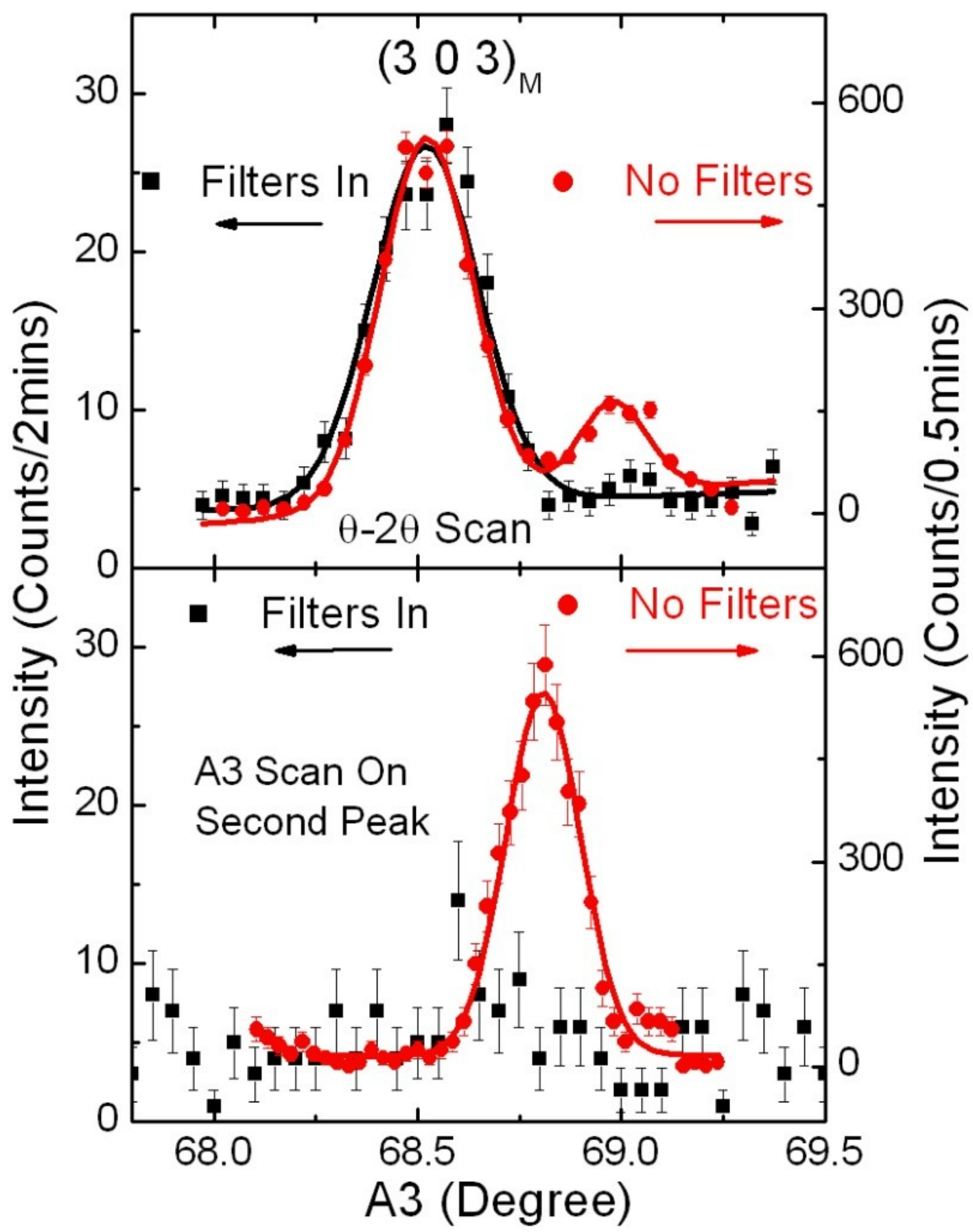

Figure 3 (color online) **a**) Rocking curves of the (3, 0, 3) magnetic Bragg peak and its comparison with structural Bragg peaks obtained from λ/2 of the nuclear (6, 0, 6) and (0, 6, 6) reflections. **b**) Identical rocking curve for (0, 3, 3) magnetic peak position showing no magnetic scattering. This provides definitive evidence that the AFM order occurs along the *a*-axis direction.

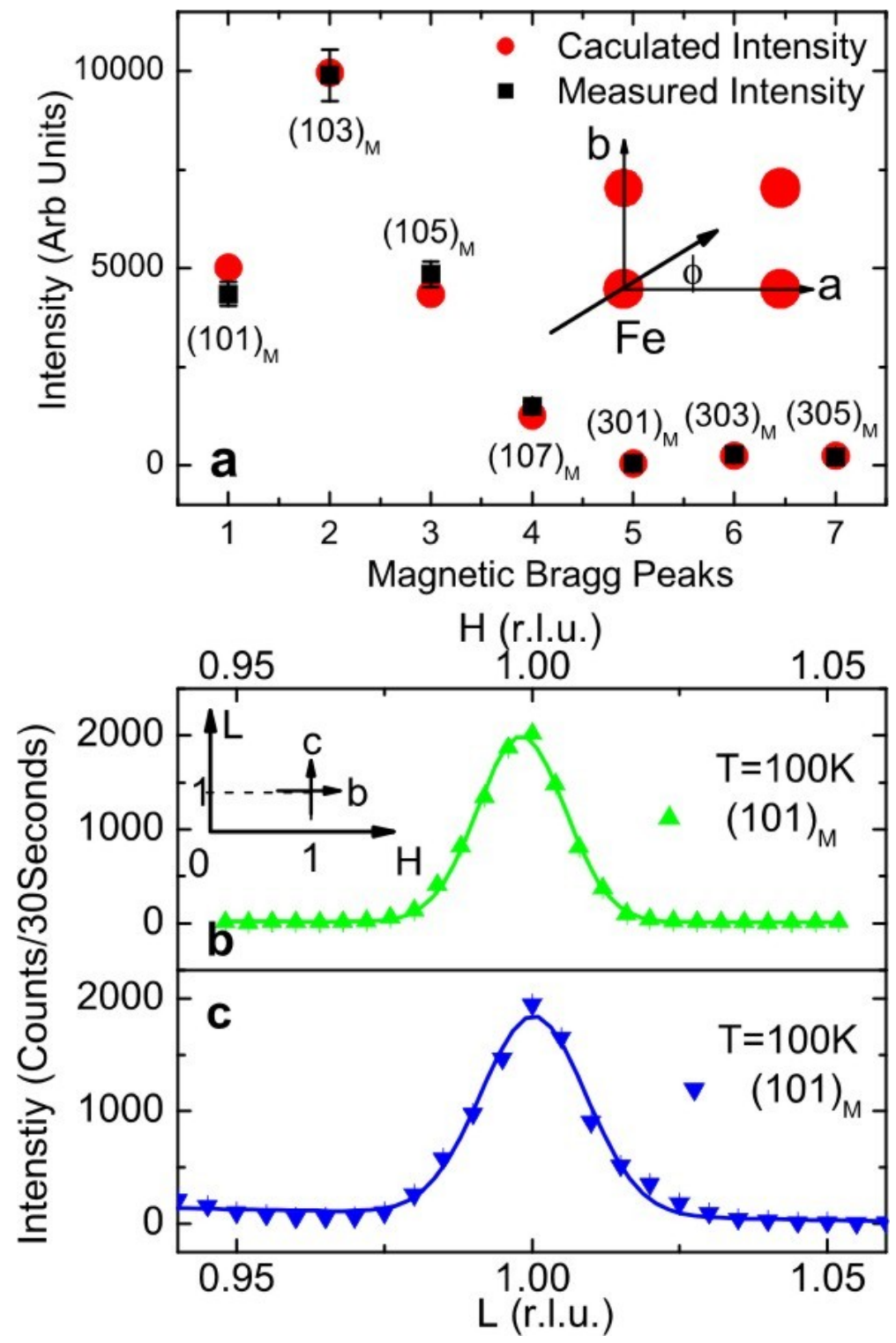

Figure 4. (color online) **a**) Calculated and observed integrated magnetic Bragg peak intensities for Fe spin direction along the *a*-axis. The agreement is excellent, demonstrating that the moment direction is along *a*. ϕ is the angle between Fe spin direction and *a*-axis, which was found to be close to zero for the best fit of the experimental data. **b**,**c**) resolution-limited [*H*, 0, 1] and [1, 0, *L*] scans through the (1, 0, 1) magnetic Bragg peak, indicating that the order in long range in nature with a minimum correlation length of 330 Å.